\documentstyle[aps,prl,floats,graphicx]{revtex}
\tighten

\newcommand\eq[1]{Eq.~(\ref{#1})}

\begin{document}

% UNCOMMENT FOR TWO-COLUMN FORMAT
 \draft \twocolumn[\hsize\textwidth\columnwidth\hsize\csname@twocolumnfalse\endcsname

\preprint{astro-ph/0107089}

\title{Correlated perturbations from inflation and the cosmic
  microwave background} 

% XXX attach two addresses to Chris
\author{Luca~Amendola$^1$, Christopher~Gordon$^{2,3}$, David~Wands$^3$ and
  Misao~Sasaki$^4$} 

\address{$^1$ Osservatorio Astronomico di Roma, Via Frascati, 00040
Monteporzio Catone, Roma}

%CCC Updated address.
\address{$^2$ DAMTP, CMS, Cambridge University, CB3 0WA, United Kingdom}

% XXX new name for Portsmouth!
\address{$^3$ Institute of Cosmology and Gravitation, University of
Portsmouth, Portsmouth~PO1~2EG, United Kingdom}

\address{$^4$ Department of Earth and Space Science, Graduate School
  of Science, Osaka University, Toyonaka 560-0043, Japan}

% \date{\today {}}
\maketitle

\begin{abstract}
  We compare the latest cosmic microwave background data with
  theoretical predictions including correlated adiabatic and CDM
  isocurvature perturbations with a simple power-law dependence.  We
  find that there is a degeneracy between the amplitude of correlated
  isocurvature perturbations and the spectral tilt.  A negative (red)
  tilt is found to be compatible with a larger isocurvature
  contribution. Estimates of the baryon and CDM densities are
  found to be almost independent of the isocurvature amplitude.  The
  main result is that current microwave background data do not exclude
  a dominant contribution from CDM isocurvature fluctuations on large
  scales.
% XXX cut 
%, and marginally favour a significant fraction.
\end{abstract}

\pacs{98.80.Cq \hfill PU-RCG-01/21, OU-TAP-164 \qquad astro-ph/0107089
  v3*}

% UNCOMMENT FOR TWO-COLUMN FORMAT
\vspace{0.5cm} 
]

%%%%%%%%%%%%%%%%%%%%%%%%%%%%%%%%%%%%%%%%%%%%%%%%%%%%%%%%%%%%%%%%%%%%

%\section{Introduction}

Increasingly accurate measurements of temperature anisotropies in the
cosmic microwave background sky offer the prospect of precise
determinations of both cosmological parameters and the nature of the
primordial perturbation spectra.
The recent Boomerang \cite{net}, DASI \cite{halverson} and
Maxima \cite{lee} data have shown evidence for 
three peaks in the cosmic microwave background (CMB)
temperature anisotropy power spectrum as expected 
in inflationary scenarios.
In this context the CMB data support the current `concordance' model
based on a spatially flat Friedmann-Robertson-Walker universe
dominated by cold dark matter and a cosmological constant \cite{tegmark}.
In addition, the CMB data no longer shows any signs of
being in conflict with the big bang nucleosynthesis data
\cite{adiablikli}. 

In the studies which have estimated the cosmological and primordial
parameters with these new data sets, only the case of purely adiabatic
perturbations has been considered so far. 
That is, the perturbation in the relative number densities
%,
%%%CCCCCC I think the delta n / n is not needed and is confusing
%$\delta n/ n$, 
of different particle species  is taken
to be zero.  
Although this
assumption is justified for perturbations originating from single
field inflationary models, it does not necessarily follow when there
is more than one field present during inflation
\cite{multiinf,LangloisI,GWBM,Hwang,bartolo}. Other possible
primordial modes are isocurvature \cite{efstathiou,bmt1} (also
referred to as ``entropy'') modes in which the particle ratios are
perturbed but the total energy density is unperturbed in the comoving
gauge. 

Most previous studies have examined the extent to which a
statistically independent isocurvature contribution to the primordial
perturbations may be constrained by CMB and large-scale structure data
\cite{indep,enqvistBM}.  
It has recently been shown that multi-field inflationary models in
general produce correlated adiabatic and isocurvature perturbations
\cite{LangloisI,GWBM,Hwang,bartolo}.  These correlations can
dramatically change the observational effect of adding isocurvature
perturbations \cite{langlois2,bmt1}.
Up until now, only the case of scale-invariant correlated adiabatic
and entropy perturbations has been considered. Trotta {\em et al.\/}
\cite{trotta} found (with an earlier CMB dataset) 
that in this case the cold dark matter (CDM) isocurvature mode
was likely to be very small if not entirely absent, though they did
find that a neutrino isocurvature mode contribution \cite{bmt1} was
not ruled out.
%
% LLLL "to which extent..."
% whether a correlated CDM isocurvature mode
% is better favoured by the recent CMB data when a tilted power law
% spectrum is allowed.
In this letter we examine to 
%%CCCCCC
%which
what
 extent a correlated
CDM isocurvature mode
is consistent with the recent CMB data when a tilted power law
spectrum is allowed.

%\section{Theory}

Non-adiabatic perturbations are produced during a period of slow-roll
inflation in the presence of two or more light scalar fields, whose
effective masses are less than the Hubble rate. On sub-horizon scales,
fluctuations remain in their vacuum state so that when fluctuations 
reach the horizon scale their amplitude is given by
%\begin{equation}
\( \label{Hover2pi}
\hat{\delta\phi}_{i*} \simeq \left( {H_* / 2\pi} \right) \hat{a}_i
\)
%\end{equation}
where the subscript $*$ denotes horizon-crossing and $\hat{a}_i$ are
independent normalised Gaussian random variables, obeying $\langle
\hat{a}_i \hat{a}_j \rangle=\delta_{ij}$. 
The total comoving curvature and entropy perturbation at any time
during two-field inflation can quite generally be given in terms of
the field perturbations, along and orthogonal to the background
trajectory, as~\cite{GWBM}
% DW: LEFT THESE EQUATIONS IN FOR NOW. BUT WE COULD TAKE THEM OUT
% LATER IF WE NEED TO FIND SPACE, AS THE PRECEDING DESCRIPTION IN
% WORDS COULD SUFFICE. 
\begin{eqnarray}
\label{RandS}
\hat{\cal R} & \propto & \cos\theta\,  \hat{\delta\phi}_1
 + \sin\theta\,  \hat{\delta\phi}_2
\,,\\
\hat{\cal S} & \propto & -\sin\theta\,  \hat{\delta\phi}_1
 + \cos\theta\,  \hat{\delta\phi}_2
\,,
\end{eqnarray}
where $\theta$ is the angle of the inflaton trajectory in field
space.
Although the curvature and entropy perturbations are uncorrelated at
horizon-crossing, any change in the angle of the trajectory,
$\theta$, will begin to introduce correlations~\cite{GWBM}.
Further correlations may be introduced by the model dependent dynamics
when inflation ends and the fields' energy is transformed into
radiation and/or dark matter.  The comoving curvature perturbation,
${\cal R}_{\rm rad}$,
on large-scales during the radiation-dominated era is
related to the conformal Newtonian metric perturbation, $\Phi$, by
${\cal R}_{\rm rad}=3\Phi/2$.  The isocurvature perturbation is ${\cal
  S}_{\rm rad}=\delta\rho_{\rm cdm}/\rho_{\rm
  cdm}-(3/4)\delta\rho_\gamma/\rho_\gamma$ and remains constant on
large scales until it re-enters the horizon.
On large scales the CMB temperature perturbation can be expressed in
terms of the primordial perturbations \cite{LangloisI}
\begin{equation}
  \label{eq:largescales}
  \frac{\hat{\delta T}}{T} \approx \frac{1}{5} \left( \hat{{\cal
  R}}_{\rm rad} - 2\hat{{\cal S}}_{\rm rad} \right) \,. 
\end{equation}

The general transformation of linear curvature and entropy
perturbations from horizon-crossing during inflation to the beginning
of the radiation era will be of the form
\begin{equation}
\label{transform}
\left( 
\begin{array}{c}
\hat{\cal R}_{{\rm rad}} \\ 
\hat{\cal S}_{{\rm rad}}
\end{array}
\right) = 
\left( 
\begin{array}{cc}
1 & T_{{\cal R}{\cal S}} \\ 
0 & T_{{\cal S}{\cal S}}
\end{array}
\right) 
\left( 
\begin{array}{c}
\hat{\cal R}_* \\ 
\hat{\cal S}_*
\end{array}
\right) \,,
\end{equation}
Two of the matrix coefficients, $T_{{\cal R}{\cal R}}=1$ and $T_{{\cal
    S}{\cal R}}=0$, are determined by the physical requirement that
the curvature perturbation is conserved for purely adiabatic
perturbations and that adiabatic perturbations cannot source entropy
perturbations on large scales~\cite{WMLL}.
The remaining terms will be model dependent.
If the fields and their decay products completely thermalize after
inflation then $T_{{\cal S}{\cal S}}=0$ and there can be no entropy
perturbation if all species are in thermal equilibrium characterised
by a single temperature, $T$. This means that it is unlikely that a
neutrino isocurvature perturbation could be produced by inflation
unless the reheat temperature is close to that at neutrino
decoupling shortly before primordial nucleosynthesis takes place. On
the other hand, a cold dark matter species could remain decoupled at
temperatures close to, or above, the supersymmetry breaking scale
yielding $T_{{\cal S}{\cal S}}$.
The simplest assumption being that one of the fields can itself be
identified with the cold dark matter~\cite{LangloisI}.

The slow evolution (relative to the Hubble rate) of light fields after
horizon-crossing translates into a weak scale dependence of both the
initial amplitude of the perturbations at horizon crossing, and the
transfer coefficients $T_{{\cal R}{\cal S}}$ and $T_{{\cal S}{\cal
    S}}$.  Parameterising each of these by simple power-laws over the
scales of interest, requires three power-laws to
describe the scale-dependence in the most general adiabatic and
isocurvature perturbations,
\begin{eqnarray}
\hat{\cal R}_{\rm rad} &=& A_r k^{n_1} \hat{a}_r + A_s k^{n_3} \hat{a}_s
 \,, \label{R} \\
\hat{\cal S}_{\rm rad} &=& B k^{n_2} \hat{a}_s \,. \label{S}    
\end{eqnarray}
The generic power-law spectrum of adiabatic perturbations from single
field inflation can be described by two parameters, the amplitude and
tilt, $A$ and $n$. Uncorrelated isocurvature perturbations require a
further two parameters, whereas we now have in general six parameters.
The dimensionless cross-correlation
 \begin{equation}
 \cos\Delta = 
 \frac{\langle{\cal R}_{\rm rad}{\cal S}_{\rm rad}\rangle}
 {(\langle{\cal R}_{\rm rad}^2\rangle \langle{\cal S}_{\rm
     rad}^2\rangle)^{1/2}}
 =  \frac{ {\rm sign}(B)\ A_s k^{n_3}}
           {\sqrt{A_r^2 k^{2n_1} + A_s^2 k^{2n_3}} } 
% \,,
\end{equation}
is in general scale-dependent.

We will investigate in this letter the restricted case where all the
spectra share the same spectral index and hence 
$\Delta$ is scale-independent.  This might naturally arise in the case of
almost massless fields where the scale-dependence of the field
perturbations is primarily due to the decrease of the Hubble rate
during inflation,  which is common to
both perturbations and yields $n_i<0$.
In the following analysis
we also allow $n_i>0$, but we shall see that blue power spectra of
this type are not favoured by the data.

We then have four parameters, $A=\sqrt{A_r^2+A_s^2}$, $B$, $\Delta$,
and $n$ describing the effect of correlated perturbations, 
where $n=1+2n_i$ is defined to coincide with the standard definition
of the spectral index for adiabatic perturbations.  
We leave an investigation of the full six parameters for future
work.

By defining the entropy-to-adiabatic
ratio \( B^{*}=B/A \) the parameter \( A \) becomes an overall amplitude
that can be marginalized  analytically (see below). In the following,
to simplify notation, we write \( A=1 \) and drop the star from \(
B^{*} \).
We limit the analysis to \( B>0 \) and \( 0<\Delta <\pi  \), since there
is complete symmetry under \( \Delta \rightarrow -\Delta  \) and
under \( (B\rightarrow -B,\Delta \rightarrow \pi -\Delta ) \). Further,
we allow three background cosmological parameters to vary, 
\(
\omega _{b}\equiv \Omega _{b}h^{2}\, \, ,\quad \omega _{c}\equiv
\Omega _{cdm}h^{2}\, \, ,\) and \( \Omega _{\Lambda } 
\)
where $\Omega_{b,cdm,\Lambda}$ is the density parameter for baryons, CDM and
the cosmological constant, respectively. Since we assume spatial
flatness, the Hubble constant
is \(
h^2={\frac{\omega _{c}+\omega _{b}}{1-\Omega _{\Lambda }}}
\).
Our aim is therefore to constrain the six parameters 
\[
\alpha_i \equiv \left\{
B,\Delta ,n, \omega _{b},\omega _{c},\Omega _{\Lambda } \right\} \,,
\]
by comparison with CMB observations.  We consider the COBE data
analysed in \cite{bon}, and the recent high-resolution Boomerang
\cite{net},
% LLLL
Maxima \cite{lee} and DASI data \cite{halverson}.  In order to
concentrate on the role of the primordial spectra (and limit the
numerical computation required) we will fix the neutrino masses (zero)
and spatial curvature (zero). We will also neglect any contribution
from tensor (gravitational wave) perturbations.

We 
% XXXX Ref.C 
% will
use a CMBFAST code \cite{sel} modified in order to allow
correlated perturbations to calculate the expected CMB angular power
spectrum, $C_l$, for all parameter values. (Our $C_l$ is defined as
$C_l=l(l+1)C^*_l/(2\pi)$ 
where $C^*_l$ is the square of the multipole amplitude). The
computations required 
can be considerably reduced by expressing the spectrum for a generic
value of \( B \) and \( \Delta \) as a function of the spectra for
other values. Let us denote the purely adiabatic and isocurvature
spectra when \( 
B=1 \) as \( [ C_{l}]_{\rm ad} \) and \( [C_{l}]_{\rm iso} \)
respectively, and the correlation term
for totally correlated perturbations \( B=1,\Delta=0 \) as \(
[C_{l}]_{\rm corr} \) . Then we can write the generic spectrum
for arbitrary $B$ and $\Delta$ as
\begin{equation}
C_{l} = [C_{l}]_{\rm ad} + 
B^{2}[C_{l}]_{\rm iso} + 2B\cos\Delta [C_{l}]_{\rm corr} 
\label{total}
\end{equation}
We can obtain $[C_{l}]_{\rm corr}$ from \eq{total} and using
any $B\cos\Delta\neq0$.  The library spectra $[C_{l}]_{\rm ad}$ and
$[C_{l}]_{\rm iso}$ and $[C_{l}]_{\rm corr}$ can then be used to
evaluate $C_{l}$ for any $B$ and $\Delta$.
A different set of library spectra will be needed for each set of cosmological
parameters.
When $n_1\not=n_3$ then $\Delta$ is not generally scale independent
and so it would be necessary to evaluate the shape of the
cross-correlation spectra $[C_{l}]_{\rm corr}$ for each
form of $\Delta(k)$, but one can always perform the scaling with
respect to $B$ analytically.

% LLLL: tau marginalization
The remaining input parameters requested by the CMBFAST code are set
as follows: \( T_{cmb}=2.726K, \) \( Y_{He}=0.24,N_{\nu}=3.04 \). All
our likelihood functions below are obtained marginalizing over
$\tau_c$, the optical depth to Thomson scattering, in the range
(0,0.2) (larger $\tau_c$ have a very small likelihood).
We did not include the cross-correlation between band powers because
it is not available, but it should be less than 10\% according to
\cite{net}.  An offset log-normal approximation to the band-power
likelihood has been advocated by \cite{bon} and adopted by
\cite{net,lee}, but the quantities necessary for its evaluation are
not available.  Since the offset log-normal reduces to a log-normal in
the limit of small noise we evaluated the log-normal likelihood
% LLLL to save  space
%\begin{equation}
%-2\log L(\alpha _{j})=\sum _{i}\frac{\left[ Z_{\ell ,t}(\ell
% _{i};\alpha _{j})-Z_{\ell ,d}(\ell _{i})\right] ^{2}}{\sigma _{\ell
% }^{2}} 
%\end{equation}
\begin{equation}
-2\log L(\alpha _{j})=\sum _{i}\left[ Z_{\ell ,t}(\ell
 _{i};\alpha _{j})-Z_{\ell ,d}(\ell _{i})\right] ^{2}\sigma _{\ell
 }^{-2} 
\end{equation}
where \( Z_{\ell }\equiv \log \hat{C}_{\ell } \), the subscripts
\( t \) and \( d \) refer to the theoretical quantity and to the
real data, \( \hat{C}_{\ell } \) are the spectra binned over some
interval of multipoles centered on \( \ell _{i} \), \( \sigma _{\ell } \)
are the experimental errors on \( Z_{\ell ,d} \), and the parameters
are denoted collectively as \( \alpha _{j} \).

The overall amplitude parameter \( A \) can be integrated out analytically
using a logarithmic measure \( d\log A \) in the likelihood.
% XXXX Ref.C(1)
Analogously, we can marginalise over the relative calibration
%CCCCC add in DASI
uncertainty of the Boomerang, Maxima and DASI data (see \cite{net,lee}),
by an analytic integration
to obtain the final likelihood function that we discuss in the following.
We neglected beam and pointing errors, but we checked that the results
do not change significantly even increasing the calibration errors by 50\%.
We assume a linear integration measure for all the other parameters.

%CCCCC add in DASI
In order to compare with the Boomerang, Maxima and DASI analyses we assume
uniform priors as in \cite{net}, with the parameters confined in the
range \( B\in (0,3), \) \( \quad \Delta \in (0,\pi ),\quad \)\ \( n\in
(0.6,1.4), \) \( \quad \omega _{b}\in (0.0025,0.08),\quad \)\ \(
\omega _{c}\in (0.05,0.4)\), \( \Omega _{\Lambda }\in (0,0.9) \). As
extra priors, the value of \( h \) is confined in the range \(
(0.45,0.9) \) and the universe age is limited to \( >10 \) Gyr as in
\cite{net}. A grid of \( \sim 10,000 \) multipole CMB spectra is used
as a database over which we interpolate to produce the likelihood
function.

\begin{figure}[ht]

{\centering \includegraphics{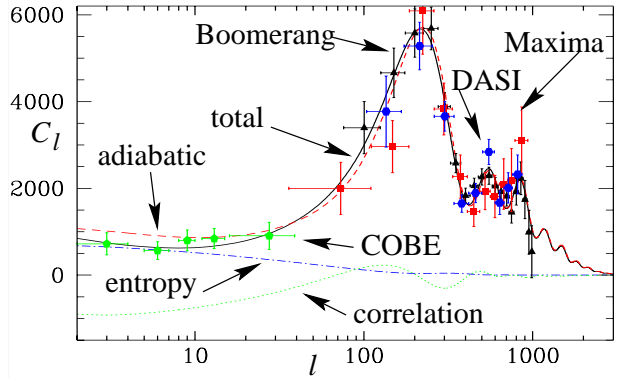}}
% XXXXX Final caption including some info moved from text!
\caption{Best-fit spectrum (solid line) and the component spectra,
  shown with the data with one sigma error bars, using
  maximum-likelihood normalisation and unadjusted relative
  calibration.}
\end{figure}

Figure~1 shows one of the best cases in our database,
corresponding to 
% XXXXX
$(B,\Delta,n,\omega _{b},\omega _{c},\Omega _{\Lambda
  })=(0.63,\pi/4,0.9,0.0225,0.1,0.7)$. 
%CCCCC change best case
%=(0.3,0.8,0.02,0.1,0.7)\) and $\Delta=0$. 
% XXXXX following info moved to caption
% %CCCC
% The data using the unadjusted calibration and maximum likelihood
% normalisation for the best fit spectrum is also displayed.
% %CCCCC
The adiabatic  $ ([ C_{l}]_{\rm ad} )$, entropy
($B^2[C_l]_{\rm iso}$) and correlated $(2B\cos\Delta[C_{l}]_{\rm corr})$
components are shown.
% XXXXX Minor rewording
The primary effect of adding a positively correlated component is to
{\em reduce} the height of the low-$l$ plateau relative to the
acoustic peaks \cite{langlois2}. This is in contrast to the
uncorrelated case where the addition of entropy perturbations
{\em increases} the plateau height relative to the peaks.
%
% XXXX Ref.C(4)
Isocurvature perturbations only have a significant effect on
intermediate angular scales for strongly blue-tilted spectra. 
They have a minimal effect on the peak structure for
% XXXXX
$n<1$.
Thus we find a near-degeneracy between \( B \) and \( n \) when \( \Delta
=0 \): the effect of adding maximally correlated isocurvature
perturbations mimics an increase in the primordial slope. This makes
clear the importance of varying \( n \) when studying correlated
isocurvature perturbations: a lower \( n \) allows a larger \( B \) to
be consistent with the CMB data. 

% XXXXX
This near-degeneracy is broken due to the effect of $n$ on the slope
at low-$l$.
In Fig. 2 we plot the likelihood for $B$ and $ {\rm cos}\Delta$,
having marginalized over the other parameters. The plot shows that the
marginalized likelihood peak occurs for $B=0.4, {\rm cos}\Delta=0.7$,
although the pure adiabatic case $B=0$ is well within one sigma. It is
remarkable that when a non-zero correlation is allowed, quite large
values of $B$ become acceptable, up to $B=1.5$ (to 95\% c.l.)  when
${\rm cos}\Delta\approx 0.8$. Anti-correlation, on the other hand,
reduces the range of $B$.
% XXXXX
We also show the likelihood contours possible in a future Planck-like
experiment with zero calibration uncertainty and accuracy
limited only by cosmic variance for $l<1000$. This shows that future
CMB data alone could detect a finite isocurvature contribution around the
current peak of likelihood. 
% XXXXX DELETE
%In the same panel we show as a long dashed line the likelihood
%assuming experimental errors reduced to one third, a precision within
%reach of the forthcoming satellite experiments: the curve shows that
%this level of precision would allow the detection of a finite
%isocurvature contribution.  
%Equally interesting, in panel \emph{b} we
%see that the likelihood of the correlation \({\rm cos} \Delta \) peaks
%near 0.6 ($\Delta \approx 60^0$)), but has a not negligible
%probability everywhere in its domain.

We found that the contour lines of the cosmological parameters \(
\omega_b \) and \( \omega_c \) are almost parallel to \( B \) for \(
B<1 \). This means that the isocurvature perturbations do not alter
significantly the best estimates for these cosmological parameters.
On the other hand, increasing \( B \) enlarges the region of
confidence for \( \Omega _{\Lambda } \) and of \( n \) toward smaller
values.

% LLL I reduced the size because now is a single plot
\begin{figure}[ht]
%%%CG fixed bounding box 
{\centering \includegraphics[width=5.6cm
%CCCCC for new fig 2
%bb=101 447 419 755]{amendola2.eps} \par} 
%   bb=92 507 361 757]{amendola2.eps} \par} 
% XXXXX CHANGED TO REDUCE SIZE AGAIN
% ,bb=92 457 361 757
,bb=87 585 285 765]
{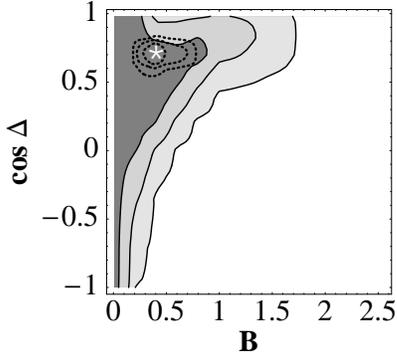} \par} 
\caption{Contour plot of the
  two-dimensional likelihood for $B$ and $ {\rm cos}\Delta$. The
  contours enclose 40\%, 86\% and 99\% of the likelihood and the star
  marks the peak.
%  See text for explanation.
}
\end{figure}

% LLLL: new fig. 3
Figure 3 summarizes our results:
we plot the one-dimensional likelihood functions obtained by
marginalizing all the remaining parameters. 
Panel \emph{a} shows that the contribution of
isocurvature perturbations can be as large as the adiabatic
perturbations, or even larger: we find that \( B<1.3 \) to 95\% c.l..
In contrast,  uncorrelated isocurvature perturbations  
 cannot exceed \( B<0.5 \) to the same c.l.. 
% 
% XXXX Ref.B
% It is
% intriguing to observe that the likelihood of \( B \) peaks around 0.3:
% that is, a non-zero contribution of isocurvature perturbations is {\em
% more} likely than a vanishing contribution.  
The likelihood functions for \( n \) and \( \Omega _{\Lambda }
\) extend toward smaller values, as anticipated, while the CDM and the
baryon density estimates remain quite unaffected. The average values
are \( n=0.94\pm 0.1\), \( \omega_b=0.023\pm 0.004\), \(
\omega_c=0.1\pm 0.03\), \( \Omega_\Lambda=0.72\pm 0.11\), .

\begin{figure}[ht]
%%%CG fixed bounding box 
{\centering 
%CCCCC change bb for new fig.
%\includegraphics[width=8.6cm, bb=69 405 473 763]{amendola3.eps} \par} 
\includegraphics[width=8.6cm, bb=110 388 448 741]{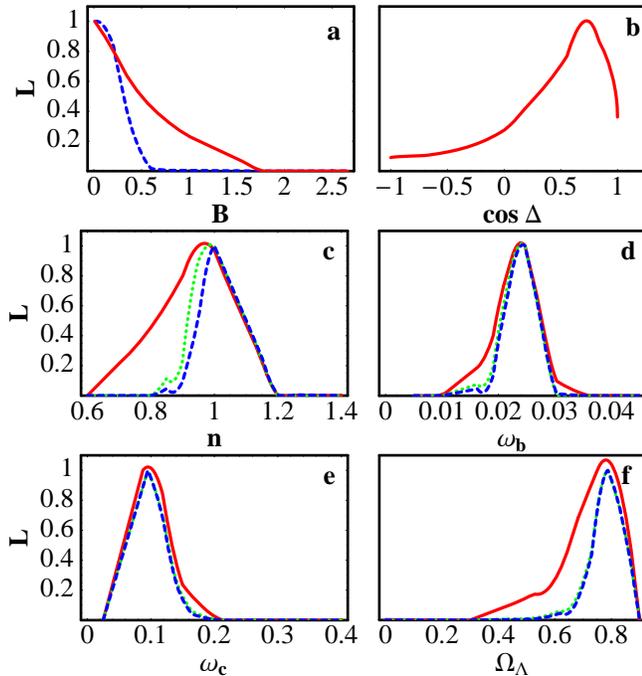} \par} 
\caption{One-dimensional likelihood functions in arbitrary units. 
  Green (light) 
%%%CCCCCC
%solid
dotted lines for the purely adiabatic models (\( B=0 \));
  blue short-dashed lines for uncorrelated fluctuations (\( \cos\Delta
  =0 \)); red (dark) solid lines for correlated fluctuations. See
  text for further explanation.} 
\end{figure}

By contrast, Enqvist {\em et al} \cite{enqvistBM} found that
a large uncorrelated isocurvature contribution is only consistent with
blue tilted slopes. The reason for this difference is that
correlations can cause the acoustic peak height to increase relative
to the Sachs Wolfe plateau (see Fig.\ 1) unlike the case of
independent perturbations where the relative height always decreases.
Trotta {\em et al} \cite{trotta} found that the CMB data was not
consistent with a significant CDM isocurvature contribution because
they restricted the primordial slope, $n$, to be unity. 
% LLLL deleted next sentences. 
% As can be seen
% from Fig.\ 2 our $n=1$ likelihood contours also indicate a very low
% isocurvature contribution. 
%
% LLLL deleted next sentence. We say the same thing a few lines below
%
% But when $n$ is allowed to be less than one
% the isocurvature contribution can be even larger than the adiabatic
% contribution.

As can be seen from Fig. 3 our estimates of $\omega_b$ and
$\omega_c$ are virtually unaffected by the addition of correlated CDM
isocurvature perturbations. Thus, in our model, the nature of the
isocurvature component can be investigated almost independently of the
composition of the matter component.

%CCCCC broaden to include DASI and COBE
The main conclusion of the present work is that 
the current CMB data
%COBE, Boomerang, Maxima and DASI are 
is
consistent with a large correlated CDM isocurvature perturbation
contribution when the spectral slopes is allowed  a tilt to the red ($n<1$).
The higher precision of future satellite data has the potential to
detect the isocurvature contribution, if any, thereby showing that
inflation was not a single-field process.

\vspace{0.3cm}

The authors are grateful to David Langlois, Roy Maartens and Carlo
Ungarelli for useful discussions.
% 
% XXXX minimise acknowledgements to save space!
%
%LA and MS are grateful to the Portsmouth Relativity and Cosmology
%Group for their hospitality where this work was initiated, supported
%by PPARC visiting fellowships grant PPA/G/S/1999/00138. 
%DW is grateful to the Osservatorio di Roma, Monteporzio, for their
%hospitality while it was continued. 
%CCCCC
CG 
%is 
was
supported by the ORS, and DW by the Royal Society.

%%%%%%%%%%%%%%%%%%%%%%%%

%CCCC Changed more than two authors to et al.

\end{document}